\documentclass{article}
\usepackage{amsmath,amssymb,amsthm,latexsym}
\usepackage[letterpaper,text={5in,9in}]{geometry}
\title{The DFAs of Finitely Different Languages}
\date{\today}
\author{Andrew Badr \and Ian Shipman}
\newtheorem{theorem}{Theorem}
\newtheorem{lemma}[theorem]{Lemma}
\newtheorem{prop}[theorem]{Proposition}

\newtheorem{alg}[theorem]{Algorithm}

\theoremstyle{definition}
\newtheorem{defn}[theorem]{Definition}

\theoremstyle{remark}
\newtheorem{remark}[theorem]{Remark}

\def\sdiff{\bigtriangleup}
\def\empty{\varnothing}
\def\iiso{\cong_I}

\begin{document}
\maketitle
\begin{abstract}
Two languages are \emph{finitely different} if their symmetric difference is finite. We consider the DFAs of finitely different regular languages and find major structural similarities. We proceed to consider the smallest DFAs that recognize a language finitely different from some given DFA. Such \emph{f-minimal} DFAs are not unique, and this non-uniqueness is characterized. Finally, we offer a solution to the minimization problem of finding such f-minimal DFAs.
\end{abstract}
\section{Preliminaries}
A DFA is a quintuple $(Q, \Sigma, \delta, q_0, A)$ following the standard definition \cite{557657}, where $Q$ is the set of states, $\Sigma$ is the alphabet, $\delta$ is the transition function, $q_0$ is the starting state, and $A$ is the set of accepting states. \\
\\
We extend the transition function $\delta$ to words in the standard way. We only consider DFAs where all states are reachable. By default, consider $D$ and $D'$ to refer to DFAs, with $D=(Q, \Sigma, \delta, q_0, A)$ and $D'=(Q', \Sigma, \delta', q'_0, A')$, and consider $L$ and $L'$ to be their languages. Finally, if $D$ is a DFA, then $L(D)$ is the language recognized by $D$.
\section{Results}
The first subsection investigates the numerous similarities between DFAs that recognize finitely different languages. It contains the bulk of our results. The second subsection addresses a natural minimization problem -- finding f-minimal DFAs. It contains a single theorem and the sketch of an algorithm.
\subsection{Main Results}
\begin{defn}[Finitely Different Languages]
If the symmetric difference $L \sdiff L'$ is a finite set, then $L$ and $L'$ are \emph{finitely different} and we write $L \sim L'$.
\end{defn}
This paper investigates the DFAs of finitely different languages. Note that the set of regular languages is closed under finite difference: if L is regular and $L \sim L'$, then L' is regular.
\begin{defn}[Equivalence Classes] 
Finite difference is an equivalence relation. The equivalence classes of this relation are called \emph{language-classes}. In a natural way, we extend this relation to DFAs such that $D \sim D'$ if $L(D) \sim L(D')$, and each DFA is likewise a member of some (equivalence) \emph{DFA-class}. 
\end{defn}
\begin{defn}[Finite Part and Infinite Part]\label{fpip}
For any DFA $D=(Q, \Sigma, \delta, q_0, A)$, $Q$ is partitioned into two sets of states: the \emph{finite part} and the \emph{infinite part}. To aid understanding, we offer two equivalent definitions of the finite and infinite parts:
\begin{enumerate}
\item For every state $q \in Q$, consider the set $\{w{\in}\Sigma^* | \delta(q_0, w) = q\}$. If this set is finite, $q$ is in the finite part of $D$, denoted by $F(D)$. If this set is infinite, $q$ is in the infinite part of $D$, denoted by $I(D)$.
\item A state $q \in Q$ is in the infinite part iff it is either on a cycle (that is, $\exists w \in \Sigma^+ | \delta(q, w)=q$) or reachable from a state which is on a cycle.
\end{enumerate}
\end{defn}

\begin{defn}[Infinite Part Isomorphism]\label{infiso}
Two DFAs $D=(Q, \Sigma, \delta, q_0, A)$ and $D'=(Q', \Sigma, \delta', q'_0, A')$ are said to have \emph{isomorphic infinite parts}, denoted by $D \iiso D'$, if there exists a bijection $f: I(D) \to I(D')$ such that
\begin{enumerate}
\item $({\forall} q \in I(D)), q \in {A} \Leftrightarrow f(q) \in A'$ and
\item $({\forall} q \in I(D), {\forall} c \in \Sigma), f(\delta(q,c)) = \delta'(f(q),c)$. 
\end{enumerate}
\end{defn}
\begin{theorem}[Infinite Part Isomorphism]\label{main}
If $D$ and $D'$ are minimized and $D \sim D'$, then $D \iiso D'$.
\end{theorem}
\begin{proof}
Let $D$ and $D'$ be minimized DFAs whose languages ($L$ and $L'$) are finitely different. For $D$, there is some length of word above which all input strings ``end up in" the infinite part.  That is, there exists a $k$ so that $|w| > k \Rightarrow \delta(q_0,w) \in I(D)$.  Likewise for $D'$.  Furthermore, since the languages have only a finite difference, there is some length of word above which the languages are identical. Let $N$ be the maximum of these three numbers.

With each state $q \in I(D)$, we associate a \emph{representative string} $w_q$ such that $\delta(q_0,w_q)=q$ and $|w_q|>N$. Strings of sufficient length must exist, since infinitely many strings reach $q$.  Now consider the function $f:I(D) \to I(D')$ defined by $f(q) = \delta'(q_0, w_q)$. We will show that $f$ is an infinite part isomorphism. 

Let $q_1 \neq q_2 \in I(D)$ and let $w_1$ and $w_2$ be their representative strings.  Since $D$ is minimized, there is a string $t$ such that $w_1t \in L$ iff $w_2t \notin L$.  Since $|w_1|,|w_2| > N$, obviously $|w_1t|,|w_2t| > N$ and therefore $w_1t \in L'$ iff $w_2t \notin L'$ by the definition of $N$.  This means that $\delta'(q'_0,w_1t) \neq \delta'(q'_0,w_2t)$, which implies that $f(q_1) = \delta'(q'_0,w_1) \neq \delta'(q'_0,w_2) = f(q_2)$.  Hence, $f$ is an injection.  We can interchange $D$ and $D'$, and choose representative strings for $I(D')$ to obtain an injection $f':I(D') \to I(D)$. Therefore $I(D)$ and $I(D')$ have the same cardinality and $f$ is a bijection. To complete the theorem, we prove that $f$ satisfies the two conditions of Definition \ref{infiso}:
\begin{enumerate}
%%%insert figure here??
\item We use a proof by contradiction. Consider any $x \in I(D)$ and $c \in \Sigma$. Let $x' = f(x)$. Let $y = \delta(x,c)$ and $z$ be such that $f(z) = \delta'(f(x),c)$. Suppose that $f(y) \neq f(z)$. Then $y \neq z$, so there exists some distinguishing string $d$ between them. If $w_x$ and $w_z$ are representative strings for $x$ and $z$ respectively, then $w_xcd \in L$ iff $w_zd \not\in L$. But in $D'$, $w_xc$ and $w_z$ go to the same state $f(z)$, so $w_xcd \in L'$ iff $w_zd \in L'$.  We are forced to conclude that $D$ and $D'$ disagree on one of $w_xcd$ and $w_zd$, but this contradicts our choice of $N$.
	\item Let $q \in I(D)$.  Since $|w_q| > N$, $w_q \in L$ iff $w_q \in L'$.  Hence, by the definition of $f$, $q \in A$ iff $f(q) \in A'$.
\end{enumerate}
\end{proof}

\begin{prop}
The converse of Theorem \ref{main} is false.
\end{prop}
\begin{proof}
Consider the minimized DFAs for $0^*$ and $10^*$. Their infinite parts are isomorphic, but the languages differ on infinitely many strings.
\end{proof}

\begin{defn}[Induced languages]
Consider a DFA $D=(Q,\Sigma,\delta,q_0,A)$.  The \emph{language induced by} $q \in Q$ is the language recognized by the DFA $(Q,\Sigma,\delta,q,A)$. This language is denoted by $L(q)$. We extend the finite difference relation to states, where if $L(p) \sim L(q)$ then $p \sim q$, and $p$ and $q$ are members of the same \emph{state-class}.
\end{defn}

\begin{defn}[$S(D)$ and $Q_C(D)$]
For any DFA $D$, define: $S(D) = \{[L(q)]: q \in Q\}$, where $[L]$ denotes the language-class of $L$.  For any language-class $C \in S(D)$, let $Q_C(D)$ denote the set of states of $D$ inducing a language in $C$.
\end{defn}

\begin{theorem}\label{sconstant}
If $D \sim D'$, then $S(D) = S(D')$.
\end{theorem}
\begin{proof}
Suppose $S(D) \neq S(D')$, with $C \in S(D) \setminus S(D')$.  For some $q \in Q_C(D)$, let $w$ be a word such that $\delta(q_0,w)=q$. Let $q' = \delta'(q'_0, w)$.  $L(q') \notin C$, so $W = L(q) \sdiff L(q')$ is an infinite set.  Since $D$ and $D'$ disagree on any word of the form $wd$, where $d \in W$, $D \nsim D'$.
\end{proof}
\begin{prop}The converse of Theorem \ref{sconstant} is false.
\end{prop}
\begin{proof}
Consider DFAs $D$ and $D'$ where $L(D) = \{w \colon |w| \text{ is odd}\}$ and $L(D') = \{w \colon |w| \text{ is even}\}$. $S(D)=S(D')$, but the DFAs disagree on infinitely many strings.
\end{proof}

\begin{lemma}\label{ipsubset}
If $D_q$ is the induced DFA of $q \in Q$ in some DFA $D$, then $I(D_q) \subset I(D)$.
\end{lemma}
\begin{proof}
Let $w$ be a word such that $\delta(q_0,w)=q$. Then for any state $q' \in Q$, $\delta(q,w')=q' \rightarrow \delta(q_0, ww')=q'$. Therefore, if any state $q'$ can be reached from $q$ by infinitely many strings, then by prepending $w$ to those strings it is clear that $q'$ can also be reached from $q_0$ by infinitely many strings.
\end{proof}

\begin{prop}\label{sdiiso}
If $D$ and $D'$ are minimized DFAs, then $S(D) = S(D') \rightarrow D \iiso D'$.
\end{prop}
\begin{proof}
Suppose $S(D) = S(D')$. Then there must exist some state $q' \in Q'$ such that $q_0 \sim q'$, where $q_0$ is the start state of $D$. Let $D'_q$ be the induced DFA of $q'$. By Lemma \ref{ipsubset}, $I(D'_q) \subset I(D')$ hence $|I(D'_Q)| \leq |I(D')|$. Since $q_0 \sim q'$, $D \sim D'_q$, so by Theorem \ref{infiso} $D \iiso D'_q$ and $|I(D)| = I(D'_q)$. Combining the two results obtains $|I(D)| \leq |I(D')|$, and by symmetry $|I(D')| \leq |I(D)|$, so $|I(D)| = |I(D')|$. Therefore, $I(D'_q) = I(D')$ and $D \iiso D'$.
\end{proof}

\begin{prop}\label{niisosd}
The converse of Proposition \ref{sdiiso} is false.
\end{prop}
\begin{proof}
Consider the minimized DFAs for $0^*$ and $10^*$. Their infinite parts are isomorphic, but no state in the former is in the same state-class as the start state of the latter.
\end{proof}

\begin{remark} 
In the results concluding with Proposition \ref{niisosd}, we have fully articulated the relationships between finite difference, $S(D)$ equivalence, and infinite-part isomorphism. In summary, $D \sim D' \rightarrow S(D) = S(D') \rightarrow D \iiso D'$, and none of the reverse implications is true. As partitions on the set of all DFAs, each is a proper refinement of the next.
\end{remark}

\begin{defn}[f-merge] 
The \emph{f-merge} operation combines two states of a DFA, given $p, q \in Q$ with $p \sim q$ and $p \in F(D)$.  To f-merge $p$ and $q$, delete $p$ and whenever $\delta(x,c) = p$, replace the transition with $\delta(x,c) = q$.  Note that since $p \in F(D)$ it is impossible for $\delta(p,c) = p$.
\end{defn}
\begin{lemma}\label{fmfin}
The f-merge operation makes only a finite difference in a DFA's language.
\end{lemma}
\begin{proof}
Suppose we are going to apply the f-merge operation to states $p,q$ of DFA $D_1$, turning it into $D_2$.  Let $X$ be the set of words that go to $p$, and let $Z$ be the set of words $L(p) \sdiff L(q)$.  The presence in $L(D_1)$ of any word not passing through $p$ is unaffected.  Considering a word of the form $xw$ for $x \in X$ we see that unless $w \in L(p) \sdiff L(q)$, the status of $xw$ with respect to $L(D_1)$ will not change.  Hence we see that $|L(D_1) \sdiff L(D_2)| = |X * Z| = |X||Z| < \infty$ since $|X|,|Z| < \infty$.  So $D_1 \sim D_2$.
\end{proof}

\begin{defn}[f-minimal] 
$D$ is \emph{f-minimal} if for any $D'$, $D \sim D' \rightarrow |Q| \leq |Q'|$.
\end{defn}

\begin{lemma}\label{fminunique}
In an f-minimal DFA, each state in the finite part is the sole representative of its state-class. In other words, if $D$ is f-minimal with $p \in F(D)$, then $p \sim q \rightarrow p=q$.
\end{lemma}
\begin{proof}
If $p \in F(D)$, $p \sim q$, and $p \neq q$, then $p$ and $q$ can be f-merged. By Lemma \ref{fmfin}, this would result in a smaller DFA of the same DFA-class, meaning $D$ could not be f-minimal.
\end{proof}

\begin{defn}[Isomorphic Finite Part]
$D$ and $D'$ are said to have \emph{isomorphic finite parts up to acceptance} if there exists a bijective function $f{\colon} F(D){\mapsto}F(D')$ such that:
$(\forall q_x, q_y\in F(D)) (\forall{c\in\Sigma}), \delta(q_x, c)=q_y \rightarrow \delta'(f(q_x), c) = f(q_y)$.
\end{defn}

\begin{theorem}\label{isofin}
If $D$ and $D'$ are f-minimal and $D \sim D'$, then their finite parts are isomorphic up to acceptance.
\end{theorem}
\begin{proof}
First, by Theorem \ref{sconstant}, $S(D)=S(D')$. Second, since all f-minimal DFAs are minimized, $D \iiso D'$, so the state-classes represented by $I(D)$ are the same as those represented by $I(D')$. So by subtraction, the state-classes represented $F(D)$ are the same as those represented by $F(D')$. By Lemma 20, or by noting that $|Q| = |Q'|$ and $|I(D)| = |I(D')|$, we may conclude that $|F(D)| = |F(D')|$. Therefore, we construct our bijection $f: F(D) \to F(D')$ by mapping each state in $F(D)$ to the state in $F(D')$ whose induced language is in the same language-class. Consider any $p, q \in F(D)$ and $c \in \Sigma$ where $\delta(p,c)=q$. The languages of $p$ and $f(p)$ differ on only finitely many strings. Since every difference between the induced languages of $\delta(p, c)$ and $\delta'(f(p), c)$ causes a difference between the induced languages of $p$ and $f(p)$ (one that begins with $c$) we conclude that $L(\delta(p,c)) \sim L(\delta'(f(p),c))$. Hence, $f(q)=\delta'(f(p), c)$, as required.
\end{proof}

\begin{remark}[Non-uniqueness of f-minimal DFAs] 
Through the finite- and infinite-part isomorphism theorems, we have shown that there must be major structural similarities between any two f-minimal DFAs of the same DFA-class. Only two aspects have not been shown to be equal: the acceptance-values of states in the finite part and the transitions that go from a finite-part state to an infinite-part state. Indeed, both of these aspects may be altered. The acceptance values of states in the finite part can be altered arbitrarily while affecting neither DFA-class nor f-minimality. As for the finite-part to infinite-part transitions, f-minimal DFAs within a class can differ on this aspect as well. However, an argument similar to that of Theorem \ref{isofin} shows that these transitions can only swap destinations within a single state-class (i.e., when there are multiple infinite-part states in the same state-class, transitions into that state-class may permute with each other). Furthermore, such a swap will preserve both DFA-class and f-minimality, while any other swap will not, so this is the best possible result.
\end{remark}

The previous results may suggest that finite language differences originate with finite-part differences. However, they may also occur when infinite parts have multiple states in the same state-class. The final result of this section demonstrates how extreme this can be.

\begin{prop} 
For any finite set of words $W$ over an alphabet with at least two characters, there exist minimized DFAs $D$ and $D'$ with $F(D) = \empty = F(D') $ and $L(D) \sdiff L(D') = W$.
\end{prop}
\begin{proof}
Let $W$ be an arbitrary finite subset of $\Sigma^*$ for some $|\Sigma| \geq 2$. Let $n = max\{|w|: w \in W\}$. We will prove the hypothesis by construction, and $D$ and $D'$ will be identical except for the starting state. The alphabet $\Sigma$ is already determined. Now, letting $\Sigma_x$ and $\Sigma^x$ be the sets of words of length at most $n$ and exactly $x$, respectively, we set $Q = \Sigma_{n} \times \{0,1\}$. Fixing a surjection $\phi:\Sigma^{n+1} \to \{(\varepsilon, 0), (\varepsilon, 1)\}$ -- such a function must exist since $|\Sigma| \geq 2$ -- we set $\delta$ as follows:
\begin{align*}
\delta( (w,i),c ) & = (wc,i) \quad \text{ if }|w| < n, \\
\delta( (w,i),c ) & = \phi(wc) \quad \text{ if }|w| = n.
\end{align*}
Let $A= \{(w, i): i=1 \text{ and } w \in W\}$. Setting $D = (Q, \Sigma, \delta, (\varepsilon, 0), A)$ and $D'=(Q, \Sigma, \delta, (\varepsilon, 1), A)$ completes our construction. It remains to prove that $F(D) =  F(D') = \empty$ and $L(D) \sdiff L(D') = W$, and that these properties are preserved by minimization. \\
\\
To prove the first property, it suffices to show that the starting states are on a cycle. We begin with $D$. Since $\phi$ is surjective, let $w_0$ be any word with $\phi(w_0) = (\varepsilon, 0)$. Then we have $\delta((\varepsilon, 0), w_0) = \phi(w_0) = (\varepsilon, 0)$. Therefore, $(\varepsilon, 0) \in I(D)$, and state reachable from $(\varepsilon, 0)$ (that is, every state) is also in $I(D)$, $F(D) = \empty$. Since a DFA's language is unchanged by minimization, the starting state $q_0$ and $\delta(q_0, w_0)$ still induce the same language. In any minimized DFA, $L(p) = L(q) \rightarrow p=q$,  so $q_0 = \delta(q_0, w_0)$ and the starting state is still on a cycle. Therefore, $F(D) = \empty$ before and after minimization. By a symmetrical proof, the same holds for $F(D')$.\\
\\
To prove the second property, begin by considering any word $w$ with $|w| \leq n$. It should be clear that $\delta((\varepsilon,i),w) = (w, i)$. Therefore, by the definition of $A$, $w \in L(D) \sdiff L(D')$ iff $w \in W$. Continuing, for any word $w$ with $w = n+1$ we have $\delta((\varepsilon,0),w) = \delta((\varepsilon,1),w) = \phi(w)$. Since $D$ and $D'$ go to the same state on any word of length n+1, they also go to the same state on any word of length greater than n+1. Therefore, $D$ and $D'$ agree on any word $w$ if $|w| \geq n+1$, so $L(D) \sdiff L(D') = W$, as desired. Finally, since minimization does not change the language of a DFA, this property too is preserved.
\end{proof}

\subsection{Algorithm}
In this section, we address the minimization problem posed by the concept of f-minimality: given a starting DFA, how can one find an f-minimal DFA in the same DFA-class? 

\begin{theorem}[No Local Minima Under F-Merge]\label{nolocalmin}
Greedy, repeated application of the f-merge operation to any minimized initial DFA will result in an f-minimal DFA of the same DFA-equivalence class as the original.
\end{theorem}
\begin{proof}
Let $D_1$ be the original minimized DFA. Since a DFA has finitely many states, f-merge can only be applied finitely many times, as each application reduces the number of states. Let $D_1...D_n$ be the sequence of DFAs reached by applying f-merge, such that $D_{k+1}$ is the result of some single application of f-merge to $D_k$, and there is no possible way to f-merge in $D_n$. Let $D_Z$ be an f-minimal DFA in the same DFA-class as $D_1...D_n$. Suppose for contradiction that $D_Z$ has fewer states than $D_n$.  By Theorem \ref{sconstant}, $S(D_n) = S(D_Z)$.  So there must exist some class $C \in S = S(D_Z)$ such that $Q_C(D_Z)$ has fewer states than $Q_C(D_n)$. Consider the number of states from $F(D_n)$ and $I(D_n)$ in $Q_C(D_n)$. If the latter is positive, then the former must be zero, or else any finite-part state in $Q_C(D_n)$ could be f-merged with an infinite-part state, contradicting our assumption that no more f-merges could be performed in $D_n$. But by Theorem \ref{main}, $D_n \iiso D_Z$, so the number of states from $I(D_n)$ in $Q_C(D_n)$ must equal the number of states from $I(D_Z)$ in $Q_C(D_Z)$. Therefore, there can be no states from $I(D_n)$ in $C$. But by Lemma \ref{fminunique} there must be exactly one state from $F(D_n)$ in $C$. Since $D_Z$ must have at least one state in $C$ (by Theorem \ref{sconstant}), there is no way it could have fewer states in $C$ than $D_n$ does, contradicting our assumption that $D_n$ was not f-minimal.
\end{proof}

\begin{alg}[F-Minimize]
Theorem \ref{nolocalmin} immediately yields an algorithm for \emph{f-minimizing} any DFA -- that is, turning it into an f-minimal DFA in the same DFA-class. This algorithm is surely suboptimal, so we only sketch the proof. The input is a DFA $D=(Q, \Sigma, \delta, q_0, A)$.
\begin{enumerate}
\item Minimize $D$ using any minimization algorithm
\item Divide $Q$ into the finite and infinite parts
\item For each pair of states $p, q$, determine whether $p \sim q$
\item Within each state-class, f-merge any $p, q$ pair where $p \in F(D)$
\end{enumerate}
The first step is standard. The second step can be accomplished by determining for each state $q$, using either depth- or breadth-first search, the set of all states reachable from $q$, and then applying the second part of Definition \ref{fpip}. The third step can be accomplished by, for each $p$ and $q$, creating a DFA recognizing the language $L(p) \sdiff L(q)$. This is done by using the standard $Q \times Q$ cross-product construction with $D_p=(Q, \Sigma, \delta, p, A)$ and $D_q=(Q, \Sigma, \delta, q, A)$ as inputs, where state $(x,y)$ is accepting if $x \in A$ xor $y \in A$. The resultant DFA is $D_{pq}$, and $p \sim q$ if after minimization $D_{pq}$ has infinite part equal to a single non-accepting state with all transitions leading to itself. (DFAs with this property recognize finite languages, and if $L(D_{pq})$ is finite then by construction $p \sim q$.) After performing the fourth step, Theorem \ref{nolocalmin} proves that the resultant DFA will be f-minimal. Step 3 dominates the running time, as it involves the costly cross-product and minimization over all pairs of states. If $n = |Q|$, then Step 3 takes $O(n^4*log n)$ time -- $n^2$ to go through each pair of states, and $n^2 log n$ on each of those to minimize the cross-product DFA. We hope and believe that there is room for improvement on this algorithm.
\end{alg}
\bibliography{finite}
\end{document}